\documentclass[journal=jpclcd,manuscript=letter,layout=twocolumn]{achemso}

\pdfoutput=1
\usepackage[latin9]{inputenc}
\usepackage{amssymb}
\usepackage{amsmath}
\usepackage{graphicx}
\usepackage{esint}
\usepackage{color}
\usepackage{pdfpages}
\usepackage{pgffor}
\usepackage{pdfpages}

\author{Martin \v{Z}onda}
\email{martin.zonda@karlov.mff.cuni.cz}

\affiliation[UK]{Department of Condensed Matter Physics, Faculty of Mathematics and Physics, Charles University, Ke Karlovu 5, CZ-121 16 Praha 2, Czech Republic}

\alsoaffiliation[Freiburg]{Institute of Physics, Albert Ludwig University of Freiburg, Hermann-Herder-Strasse 3,
79104 Freiburg,
Germany}

\author{Oleksandr Stetsovych}

\affiliation[CzAS]{Institute of Physics, Czech Academy of Sciences, Cukrovarnick\'a 10,
CZ-162 00 Praha 6, Czech Republic}

\author{Richard Koryt\'{a}r}

\affiliation[UK]{Department of Condensed Matter Physics, Faculty of Mathematics and
Physics, Charles University, Ke Karlovu 5, CZ-121 16 Praha 2, Czech
Republic}

\author{Markus Ternes}

\affiliation[Aachen]{Institute of Physics II B, RWTH Aachen University, 52074 Aachen, Germany}
\alsoaffiliation[Julich]{Peter Gr\"unberg Institut (PGI-3), Forschungszentrum J\"ulich, Germany}

\author{Ruslan Temirov}

\affiliation[Julich]{Peter Gr\"unberg Institut (PGI-3), Forschungszentrum J\"ulich, Germany}
\alsoaffiliation[Cologne]{University of Cologne, Faculty of Mathematics and Natural Sciences, Institute of Physics II, Germany}

\author{Andrea Raccanelli}

\affiliation[JulichC]{Peter Gr\"unberg Institute (Cryo-Lab), Forschungszentrum J\"ulich, Germany}

\author{F. Stefan Tautz}

\affiliation[JulichC]{Peter Gr\"unberg Institute (PGI-3), Forschungszentrum J\"ulich, Germany}
\alsoaffiliation[Jara]{J\"ulich Aachen Research Alliance (JARA), Fundamentals of Future Information Technology, Germany}
\alsoaffiliation[Aachen2]{Institute of Physics IV A, RWTH Aachen University, Germany}

\author{Pavel Jel\'{i}nek}

\affiliation[CzAS]{Institute of Physics, Czech Academy of Sciences, Cukrovarnick\'a 10,
CZ-162 00 Praha 6, Czech Republic}

\alsoaffiliation[RCPTM]{RCPTM, Palacky University, \v{S}lechtitelu 27, 783 71, Olomouc, Czech Republic.}

\author{Tom\'a\v{s} Novotn\'y}

\affiliation[UK]{Department of Condensed Matter Physics, Faculty of Mathematics and
Physics, Charles University, Ke Karlovu 5, CZ-121 16 Praha 2, Czech
Republic}

\author{Martin \v{S}vec}
\email{svec@fzu.cz}
\affiliation[CzAS]{Institute of Physics, Czech Academy of Sciences, Cukrovarnick\'a 10,
CZ-162 00 Praha 6, Czech Republic}
\alsoaffiliation[RCPTM]{RCPTM, Palacky University, \v{S}lechtitelu 27, 783 71, Olomouc, Czech Republic.}

\title{Resolving Ambiguity of the Kondo Temperature Determination in Mechanically Tunable Single-Molecule Kondo Systems}

\abbreviations{}

\begin{document}

\begin{tocentry}
   \includegraphics[width=1.0\textwidth]{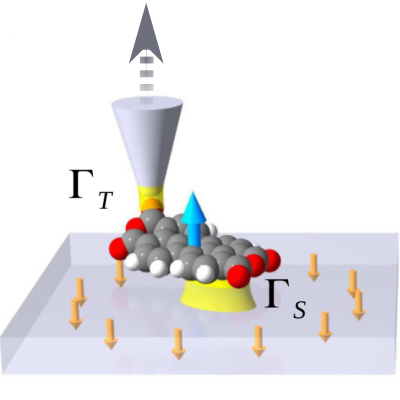}
\end{tocentry}	
 	
\begin{abstract}
Determination of the molecular Kondo temperature $T_K$ poses a challenge in most cases when the experimental temperature cannot be tuned to a sufficient extent. We show how this ambiguity can be resolved if additional control parameters are present, such as magnetic field and mechanical gating.
We record the evolution of the differential conductance by lifting an individual molecule from the metal surface with the tip of a scanning tunneling microscope. By fitting the measured conductance spectra with the single impurity Anderson model we are able to demonstrate 
that the lifting tunes the junction continuously from the strongly correlated Kondo-singlet
to the free spin $1/2$ ground state. In the crossover regime, where $T_K$ is similar to the temperature of experiment, 
the fitting yields ambiguous estimates of $T_K$ varying by an order of magnitude. 
We show that analysis of the conductance measured in two distinct external magnetic fields can be used to resolve this problem.
\end{abstract}


\maketitle

The Kondo effect is one of the most prominent many-particle phenomena in which a 
magnetic moment of an atom or molecule is screened by the itinerant electrons of the metallic substrate\cite{Hewson97}. Its highly correlated singlet ground state is  
fully characterized by the energy $k_{\mathrm{B}}T_{\mathrm{K}}$, where 
$k_{\mathrm{B}}$ is the Boltzmann constant and $T_{\mathrm{K}}$ the Kondo 
temperature. $T_{\mathrm{K}}$ depends exponentially on the strength of the 
antiferromagnetic exchange coupling between the localized moment and the 
metallic substrate. The Kondo effect is fundamental to various 
phenomena at nanoscale, ranging from the occurrence of zero-bias 
anomalies in transport measurements through quantum dots \cite{Goldhaber98, 
Parks07, Parks10}, nanowires \cite{Nygard00}, single atoms and molecular junctions \cite{evers2020advances}, and molecular magnets~\cite{Madhavan98,Yu04,Otte08a,Coronado2020,Chen2021}, to quantum critical phenomena 
\cite{Jones88,*Jones89,Lechtenberg17,Esat16}, 
and has implications for spintronics 
\cite{Bergmann15}. 

Scanning probe microscopy (SPM) allows studies of the Kondo effect in atomic and  molecular systems with unprecedented detail. 
Since the first SPM observations  of the Kondo effect on individual magnetic adatoms adsorbed on noble metal  surfaces \cite{Madhavan98,Li98a,Madhavan01,Knorr02} it has become usual to  extract $T_{\mathrm{K}}$ from the differential $dI/dV(V)$ conductance spectra measured over the Kondo impurity: 
When the experimental temperature $T_{{\rm exp}}$  is much lower than $T_{\mathrm{K}}$ ($T_{{\rm exp}} \ll T_{\mathrm{K}}$), the system is in the strongly-correlated Kondo  singlet state. 
The spectra are dominated by a zero-bias Kondo resonance whose half width at half maximum (HWHM) is directly proportional to $k_{\mathrm{B}}T_{\mathrm{K}}$  \cite{Hewson97}. 
However, when $T_{{\rm exp}} \gg T_{\mathrm{K}}$, the spin of the magnetic impurity becomes asymptotically free and the tunneling spectra are best described  by a perturbative spin-flip model, yielding temperature-broadened logarithmic singularities at zero bias \cite{appelbaum66,*appelbaum67,Zhang13,ternes15}.

The crossover between both regimes can be controlled via substrate hybridization 
\cite{Temirov:Nanotech08,Greuling11,Toher11,greuling13,Oberg13}. 
However, in spite of a significant experimental effort \cite{Parks07, Parks10,Temirov:Nanotech08,Greuling11,Toher11,greuling13,Oberg13}, a comprehensive study of the continuous 
evolution of $T_{\mathrm{K}}$ still poses a significant challenge.   
One of the reasons is the difficulty in determining the relevant energy scale in the crossover regime from 
experimental data, which can lead to a drastically incorrect estimate of 
$T_{\mathrm{K}}$ \cite{Zhang13, Gruber18}. In 
this Letter we provide both the experimental data and the recipe for their 
quantitative analysis which are together showing a  
continuous evolution from strongly correlated Kondo-regime to the free spin $1/2$ ground-state. 
Our key finding is that $T_{\mathrm{K}}$ cannot be deduced unambiguously from the  
\emph{experimental} $dI/dV$ spectra unless additional information is taken into account. We show that incipient (not fully evolved) Zeeman splitting is sufficient for this goal if two cases with distinct magnetic fields are analyzed together. 

\begin{figure}
\includegraphics[width=0.85\columnwidth]{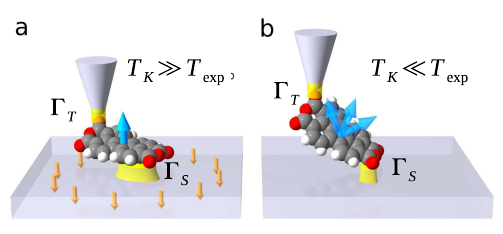} \caption{ 
Schematics of the experiment. Lifting the molecule by changing the 
tip-sample distance $z$ controls the coupling $\Gamma_S$ between the molecule 
and the substrate, and consequently, the system transforms from the strong 
coupling regime with $T_K\gg T_{\rm exp}$ (a), to the weakly-coupled spin-flip 
regime with $T_K\ll T_{\rm exp}$ (b). The coupling $\Gamma_T$ of the molecule to 
the tip remains small and mainly constant.}
\label{fig1_scheme} 
\end{figure}

We use the archetypical 3,4,9,10-perylene-tetracarboxylic-dianhydride (PTCDA) 
molecule on the Ag(111) surface as our model system. The tuning of the 
hybridization is achieved by lifting the molecule with the tip of an SPM from 
the surface \cite{Temirov:Nanotech08,Greuling11,Toher11,greuling13}. The 
variation of the tip-sample distance during the lifting process provides 
effective control over the (dominant) coupling $\Gamma_\mathrm{S}$ between the 
singly occupied molecular orbital and the substrate. It is thus possible to 
transform the system from the strongly correlated Kondo singlet to the free spin $1/2$ ground state (Fig.~\ref{fig1_scheme}).

In the lifting experiments, we record $dI/dV$ spectra as a function of 
the vertical tip position $z$, where  
$z=0$ is defined as the tip position at which the tip apex atom makes contact with 
one of the carboxylic oxygen atoms of PTCDA (Fig.~\ref{fig1_scheme}). This 
corresponds to a tip-sample distance of approximately 
$6.3$~\AA~\cite{Greuling11}. All experiments in this study are carried out at a 
base temperature of $1.2$~K. We focus on tip positions $z\gtrsim200$~pm, where a symmetric resonance at zero bias is observed in the $dI/dV$ spectra.

\begin{figure}[t]
\includegraphics[width=1\columnwidth]{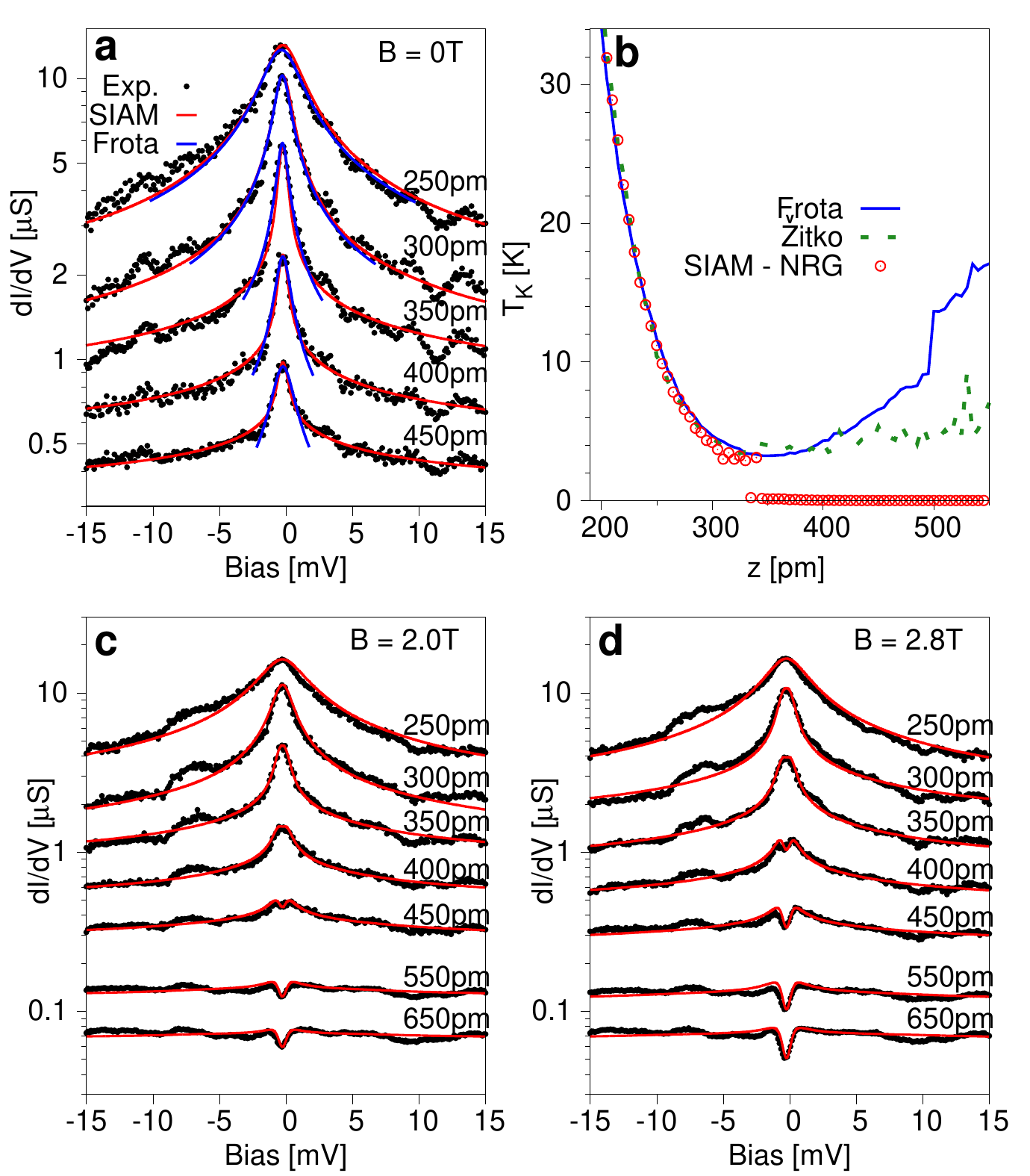} 
\caption{(a) Spectra obtained for $B=0$ at different 
tip heights  $z$ (dots) and their fits using Eq.~\eqref{eq:Frota} (blue lines) 
and NRG calculations (red lines). (b) Apparent  $T_{\mathrm{K}}$ extracted from the line shapes using different types of fits : Frota fit (full line), procedure introduced by \v{Z}itko  \cite{zitko11a}  
(dashed line) and direct fit to SIAM using NRG, for a broad range of the tip-substrate distances $z$. (c, d) 
$dI/dV$ spectra (dots) for $B=2.0$ (c) and $2.8$~T (d) and NRG-based fits 
(lines). Note the logarithmic scale on the vertical axis in panels (a), (c), and (d).}
\label{fig2_cond-Frota} 
\end{figure}

The evolution of the $dI/dV$ spectra in the range $z<200$~pm 
was previously discussed in Refs.~\cite{Temirov:Nanotech08,Greuling11,Toher11,greuling13}. In the flat adsorbed geometry ($z\ll 200$~pm) the occupancy of the lowest unoccupied molecular orbital (LUMO) is close to 2 electrons. Reducing the occupancy of the LUMO to 1 by lifting the molecule leads to the Kondo effect regime that we analyze here in detail.
Fig.~\ref{fig2_cond-Frota}a displays five 
subsequently acquired spectra at $B=0$. The broad Kondo peak at $z=250$~pm 
becomes progressively sharper as the molecule is decoupled from the substrate by 
increasing $z$, and the overall conductance is gradually decreasing.

As the first step to rationalize the data in Fig.~\ref{fig2_cond-Frota}a, we fit 
the  spectra with the standard phenomenological Frota formula (blue curves in Fig.~\ref{fig2_cond-Frota}a) 
\cite{frota86,*frota92},
\begin{equation}
\frac{dI}{dV}\left(V\right)\propto \mathcal{R}\sqrt{\frac{i\Gamma_{\mathrm{F}}}{i\Gamma_{\mathrm{F}}+eV}},\label{eq:Frota}
\end{equation}
where the Frota width $\Gamma_{\mathrm{F}}$ is related to the Kondo temperature 
by $k_{\mathrm{B}}T_{\mathrm{K}}=0.686\,\Gamma_{\mathrm{F}}$ which holds when $T_{\mathrm{K}}\gg T_{\mathrm{exp}}$~\footnote{The value 
of the numerical prefactor depends on the definition of $T_{\mathrm{K}}$. We use 
the Wilson's definition (for wide band limit) as it is common in NRG, 
perturbation theory, and Bethe Ansatz studies. A short discussion on how this 
definition relates to other commonly used ones can be found in 
Ref.~\cite{zitko11a}.}. We observe a drop in the apparent $T_{\mathrm{K}}$ from 
$\approx32$~K at $z\simeq 200$~pm to $\approx4$~K at $z\simeq 350$~pm 
(Fig.~\ref{fig2_cond-Frota}b) \footnote{Note that a simple finite-temperature 
correction (Ref.~\cite{Zhang13}, Suppl.\,Note 2) does not differ significantly 
from the zero-temperature results.}. However, the apparent $T_{\mathrm{K}}$ 
steadily rises when $z$ is increased further. 
The observed minimum is  
highly suspicious because $T_{\mathrm{K}}$ depends exponentially on 
the total coupling ($\Gamma=\Gamma_{\mathrm{T}}+\Gamma_{\mathrm{S}}$) of the 
molecule to tip ($\Gamma_{\mathrm{T}}$) and sample ($\Gamma_{\mathrm{S}}$). 
Specifically, $T_{\mathrm{K}}$ calculated for the suitable spin $1/2$ single impurity Anderson model (SIAM) reads \cite{Haldane_1978, Krishna-murth_1980}
\begin{equation}
k_{\mathrm{B}}T_{\mathrm{K}}=0.29\sqrt{\Gamma U}\exp\left(-\frac{\pi U}{8\Gamma}\right),
\label{eq:Tk}
\end{equation}
where $U$ is the intraorbital Coulomb repulsion~\cite{Haldane_1978,Krishna-murth_1980} and the particle-hole symmetry is assumed.
Because it is expected that lifting reduces $\Gamma_{\mathrm{S}}$ but keeps $\Gamma_{\mathrm{T}}$ approximately constant, any rise of $T_{\mathrm{K}}$ with increasing $z$ is counterintuitive. 
The first hint, that the problem might be in the estimation of $T_{\mathrm{K}}$ comes  
from an alternative fitting procedure proposed by \v{Z}itko 
\cite{zitko11a}. Although it 
leads to a similar behavior as Frota fitting (Fig.~\ref{fig2_cond-Frota}b) 
the increase of $T_{\mathrm{K}}$ is much less dramatic. 

Recalling that the Frota fit is bound to fail if $T_{\mathrm{K}}\lesssim T_\mathrm{exp}$
we turn to the numerical renormalization group (NRG) theory.  
Because the contact of the PTCDA molecule to the tip is electronically weak 
\cite{Temirov18}, we can assume a strongly asymmetric junction, resulting in an 
asymmetry parameter $a\equiv\Gamma_\mathrm{T}/\Gamma_\mathrm{S} \ll 1$ 
(Fig.~\ref{fig1_scheme}). This justifies the interpretation of the $dI/dV$ 
spectra as temperature-broadened equilibrium Kondo spectral functions, 
calculated as \cite{Esat_2015}
\begin{equation}\label{eq:FitFormula}
\frac{dI}{dV}(V)=-G\int_{-\infty}^{\infty}d\omega\,\pi\Gamma\,\rho(\omega,\Gamma
,T,B)f'(\omega-eV,T),
\end{equation}
where the impurity spectral function $\rho(\omega,\Gamma,T,B)$ 
of the SIAM calculated with NRG depends on the temperature $T$, magnetic field $B$, 
and total coupling $\Gamma$. Here, $f'(\omega,T)$ is the derivative of the 
Fermi-Dirac distribution with respect to $\omega$. $\Gamma_\mathrm{T}$ and 
$\Gamma_\mathrm{S}$ are assumed to be independent of $\omega$ and $B$. Hence,
\begin{equation}\label{eq:asymmetry}G\equiv\frac{2e^2}{h}\frac{4\Gamma_S\Gamma_T}{\Gamma^2}=\frac{2e^2}{h}\frac{4a}{(1+a)^2}
\end{equation}  
depends only on the coupling asymmetry $a$, while the integrand in 
Eq.~\eqref{eq:FitFormula} depends on the total coupling $\Gamma$.

The SIAM spectral functions calculated with NRG are commonly used for a qualitative comparison with experiments \cite{Bulla2008}. However, a direct least-squares fitting of experimental spectra (shown, e.g., in Fig.~\ref{fig2_cond-Frota}a,c,d) is largely avoided because of its computational complexity (for recent counterexamples see, e.g., Refs.~\cite{Jiang2018,Minamitani2015,Sturm2020}). 
Nevertheless, as we discus in detail in the Supporting Information (SI),
modern NRG packages~\cite{Ljubljana_code} make this task manageable. In our calculations we fix $T$ and $B$, and we have also found out that varying $U$ within the expected range of $0.5-1$~eV does not significantly change the fitted $T_{\mathrm{K}}$ as its influence can be mitigated by varying $\Gamma$. Therefore, our relevant fitting parameters are $G$ and $\Gamma/U$ (see 
Eq.~\eqref{eq:Tk}). In practice we precalculated large sets of spectral functions of SIAM for a dense mesh of $\Gamma/U$ values with $G=1$. These are then used to approximate the experimental $dI/dV$ curves by first fitting the optimal $G$ and then picking $\Gamma/U$ curves with the minimal sum of the least-squares residuals.

Kondo temperature obtained from the NRG fits (Fig.~\ref{fig2_cond-Frota}b) initially follows the Frota-estimated $T_{\mathrm{K}}$. However, around its minimum value at $z\approx350$~pm, the NRG fit drops below $1K$. This is in compliance with the expectation that $T_{\mathrm{K}}$ should monotonously decrease with increasing $z$. 
Nonetheless, one still has to be cautious with the interpretation of the results. Especially intriguing is the sudden discontinuous drop of $T_K$ over more than an order of magnitude. In addition, it is important to note that an analogous study of the weak coupling regime in a different system and with a different fitting procedure showed that the determination of $T_{\mathrm{K}}$ from just the zero-field data can be tricky and even erroneous \cite{Zhang13}.  
    
\begin{figure}
\includegraphics[width=1\columnwidth]{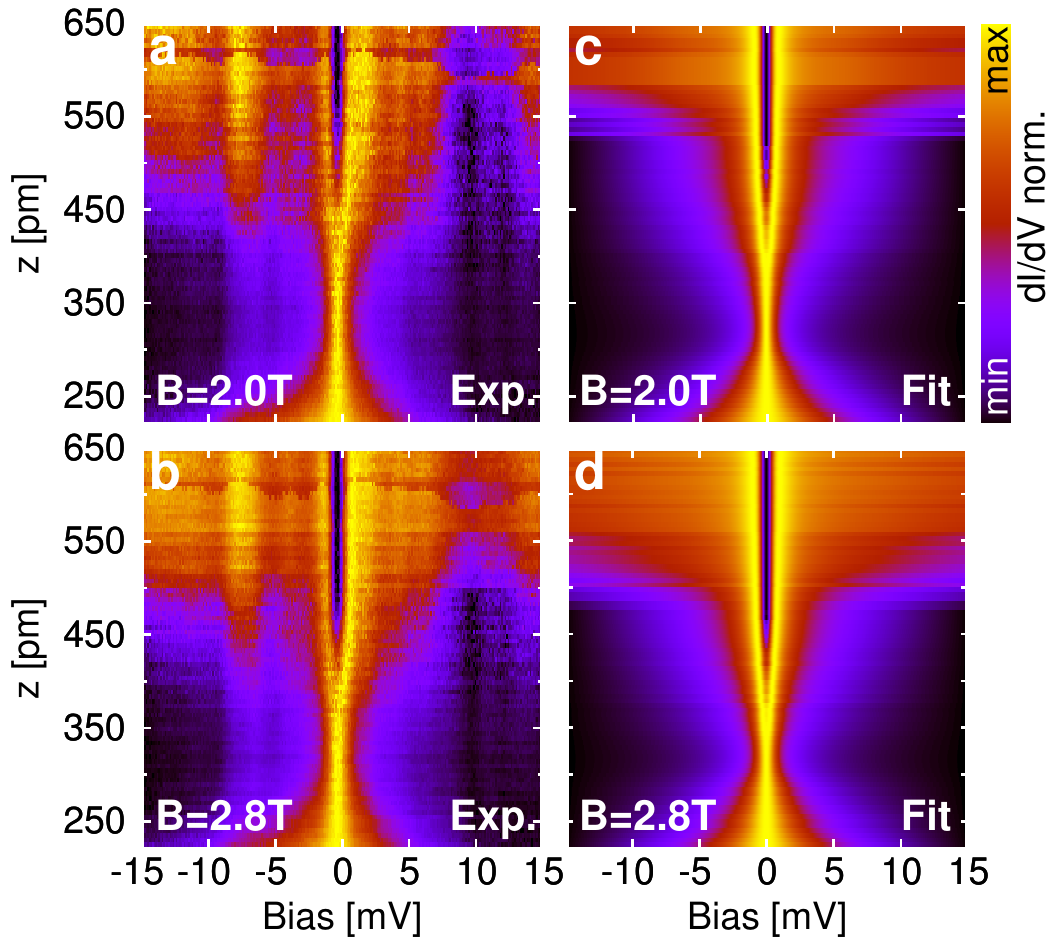} \caption{
(a, b) 
Experimental data represented by spectral intensity maps. Each measured spectrum 
(horizontal line) is normalized by subtracting its minimum value and dividing by 
its total range. (c, d) Equivalent maps obtained from the modeled spectra using 
NRG.}
\label{fig3_maps} 
\end{figure}

Therefore, in order to investigate the peculiar behavior of $T_K$ with changing $z$ and to stabilize the fitting in the (expected) weak coupling regime we repeated the lifting experiments in finite magnetic fields (Fig.~\ref{fig2_cond-Frota}c,d and Fig.~\ref{fig3_maps}a,b).  

The presence of 
the magnetic field leads to a splitting of the zero-bias conductance 
peak at sufficiently large $z$. 
After an initial increase, the splitting remains constant during further tip retraction, as seen in  Fig.~\ref{fig3_maps}a,b.

For a visible splitting of the Kondo resonance, the Zeeman energy of 
$g\mu_{\mathrm{B}}B\gtrsim2k_{\mathrm{B}}T_{\mathrm{K}}$ is needed 
\cite{costi00,zitkoPHD}. 
The fact that the splitting occurs at $z\simeq 
450$~pm for $B=2.0$~T but already at  $z\simeq 400$~pm for $2.8$~T implies that $T_{\mathrm{K}}(z=450~\mathrm{pm})< T_{\mathrm{K}}(z=400~\mathrm{pm})$. This is in a direct contradiction with the picture of a constant or even increasing 
$T_\mathrm{K}$ for $z>350$~pm as drawn from a simple Frota fit in Fig.~\ref{fig2_cond-Frota}b. 
 
The NRG fits account for both, the changing 
width of the Kondo resonance, as well as its splitting for 
$g\mu_{\mathrm{B}}B\gtrsim2k_{\mathrm{B}}T_{\mathrm{K}}$ (Fig.~\ref{fig3_maps}) \footnote{The 
experimental $dI/dV$ displays additional features that we interpret as arising 
from inelastic tunneling processes, see SI for details.}. 
From the NRG fits, two important results are principally available: a mapping $z 
\rightarrow \Gamma/U$, and, via Eq.~\eqref{eq:Tk}, the Kondo temperature as a function of the tip height $T_{\mathrm{K}}(z)$.  

\begin{figure}[htbp]
\includegraphics[width=0.99\columnwidth]{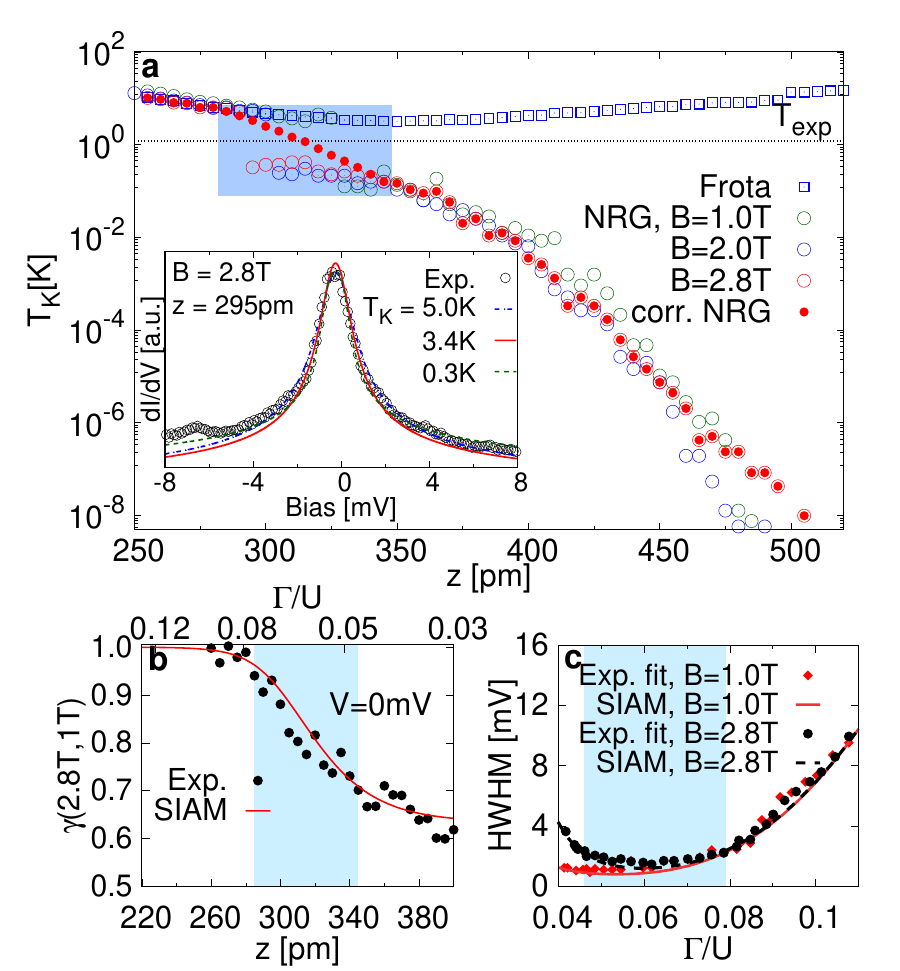} 
\caption{(a) Kondo temperature $T_{K}$ as determined by Frota approximation (blue squares) for $B=0$, direct NRG-based SIAM fitting (empty circles) for $B\neq0$ and corrected NRG fitting discussed in the text. The inset demonstrates that experimental $dI/dV$ can be in the crossover region (blue rectangle) fitted with different values of $\Gamma/U$ ($0.051$, $0.070$, $0.075$) leading to huge uncertainty in the estimation of $T_K$.
(b) Ratio $\gamma(B_1,B_2)$ at $B_1=2.8$~T and $B_2=1$~T (dots: experimental data, line: 
NRG fit). The ratio changes strongly in the transition regime (blue-shaded area) enabling 
the determination of $\Gamma/U$. The upper x-scale shows some values of $\Gamma/U$ at respective $z$. (c) The HWHM of the $B=1$~T and $2.8$~T data 
sets are almost constant in the transition regime (blue-shaded area), 
illustrating the fitting complication. 
\label{fig4_fit-NRG}}
\end{figure}

The function $T_{\mathrm{K}}(z)$ obtained from the NRG-based fitting for various $B$ drops roughly exponentially by nine orders of magnitude over the experimental 
$z$ range (Fig.~\ref{fig4_fit-NRG}a). Clearly, the huge discrepancy between the Frota fit and the NRG-based fit for 
$z\gtrsim350$~pm is a direct consequence of the fact that in the weak-coupling regime the $B=0$ zero-bias anomaly is not a Kondo resonance but a 
temperature-broadened logarithmic singularity, whose shape nevertheless resembles the Frota function. Consequently, the application of 
Eq.~\eqref{eq:Frota} leads to a dramatic overestimation of $T_\mathrm{K}$ by 
several orders of magnitude (Fig.~\ref{fig4_fit-NRG}a). In contrast, the split 
Kondo peak at sufficiently high $B$ makes the data sets in this $z$ range feature-rich, which allows a meaningful fit with NRG-based SIAM spectra.

However, the direct NRG fits at finite $B$ do not 
solve the mystery of the discontinuous drop of the $T_{\mathrm{K}}$ 
in the intermediate coupling regime, as they show the same feature.
Nevertheless, the fact that the drop happens at different $z$ for different fields $B$ and even shows some instabilities does not point to a true physical interpretation but rather again to a problem with the fitting procedure.

Indeed, while the NRG-based fitting works well for the strong coupling as well as for the split resonance in the 
weak-coupling limit, it is  problematic in the crossover regime where 
$T_{\mathrm{K}}\simeq T_{\mathrm{exp}}$ (shaded blue area in 
Fig.~\ref{fig4_fit-NRG}). 
The reason is that the width of 
the yet unsplit Kondo peak in this regime is  
not very sensitive to $T_K$ (or, $\Gamma/U$).
This produces fitting instabilities because the change of the height of the peak, which \emph{is} sensitive to $\Gamma/U$, can be mimicked by the variation of the other fitting parameter $G$. 
This is demonstrated in the inset of Fig.~\ref{fig4_fit-NRG}a
where by tuning $G$ a very good fit of the experimental data from the problematic regime can be achieved by significantly different values of $\Gamma/U$. 
As a consequence, no unique mapping $z \rightarrow \Gamma/U$ can be established in this 
range by directly comparing experimental and theoretical spectra
unless the prefactor $G$ is eliminated from the fitting.

We have found out that a convenient way to surmount this difficulty
is to assume that $G$ does not depend on $B$  
in the crossover regime and investigate the $G$ independent ratio $\gamma(B_1,B_2)=[dI/dV(0,B_1)]/[dI/dV(0,B_2)]$ between zero-bias conductances obtained 
at two significantly different magnetic field strengths $B_x$.  
In particular one with 
$g\mu_{\mathrm{B}}B>2k_{\mathrm{B}}T_{\rm exp}$, and the other with 
$g\mu_{\mathrm{B}}B<2k_{B}T_{\rm exp}$, e.g., at $B=2.8$~T and $1$~T as plotted in Fig.~\ref{fig4_fit-NRG}b.

For small $z$, we find  $\gamma\approx 1$, because here 
$T_{\mathrm{K}}>T_{\mathrm{exp}}$ and both magnetic fields are too small to affect the Kondo resonance. 
For large $z$, we find $\gamma<1$, because the 
$B$-field of $2.8$~T splits the Kondo resonance, leading to a reduced zero-bias 
conductance. In the crossover regime $\gamma$ changes 
rapidly and nonlinearly (blue shaded area in Fig.~\ref{fig4_fit-NRG}b where circles represent experimental data). Yet, there is no sign of a discontinuity in the experimental data. Aligning this rapidly 
changing curve with the same quantity calculated for SIAM by NRG as a function of 
$\Gamma/U$ (for details see SI) allows us to establish an unambiguous mapping $z \rightarrow 
\Gamma/U$ in the crossover regime (Fig.~\ref{fig4_fit-NRG}b). Given this, the correct smooth behavior of $T_\mathrm{K}(z)$ can be determined by Eq.~\eqref{eq:Tk} (full red circles in Fig.~\ref{fig4_fit-NRG}a).

Because the corrected $T_\mathrm{K}$ in the crossover regime is estimated 
using only zero-bias conductance we also provide 
a consistency test of $z \rightarrow \Gamma/U$ mapping. We refit the complete $dI/dV$ evolution using $G$ as the only remaining free parameter in the crossover regime [Fig.~\ref{fig3_maps}(c,d)]. 
We demonstrate the quality of the fit by comparing the HWHM of the experimental and NRG-calculated spectra in Fig.~\ref{fig4_fit-NRG}c. The agreement between experiment and theory is excellent throughout the crossover regime where we observe a very weak dependence of the HWHM on $\Gamma/U$. It is this weak dependence of HWHM on $\Gamma/U$ that produced an ambiguity in our initial attempt to achieve $z \rightarrow \Gamma/U$ mapping by comparing experimental and NRG-based conductance spectra. Outside the crossover regime, the HWHM rapidly rises, either because of an increasing $T_\mathrm{K}$ (right) or the incipient split of the Kondo resonance by the applied magnetic field (left).

\begin{figure}
\includegraphics[width=0.9\columnwidth]{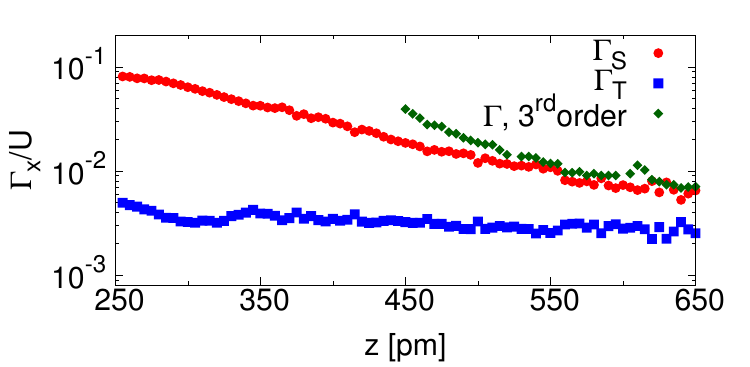} 
\caption{
Extracted coupling strengths of the molecule to the 
substrate $\Gamma_S/U$ and tip $\Gamma_T/U$, respectively, and using the third order 
perturbation theory\label{fig5_fit-ratio} showing the total $\Gamma/U$ \cite{ternes15}.}
\end{figure}

An additional benefit of the above consistency test 
is the reliable extraction of $G$ in the otherwise problematic crossover regime. 
Consequently, we can use Eq.~\eqref{eq:asymmetry} together with $\Gamma=\Gamma_\mathrm{S} + \Gamma_\mathrm{T}$ to determine the coupling constants $\Gamma_{T}/U$ and $\Gamma_{S}/U$ separately (Fig.~\ref{fig5_fit-ratio}).    
We find that across the complete $z$ range studied here, $\Gamma_\mathrm{S} \gg 
\Gamma_\mathrm{T}$. This retroactively justifies the model underlying Eq.~\eqref{eq:FitFormula}, that is, the approximation of the molecule being 
equilibrated with the substrate and the tip serving as a weak probe of the 
spectral function.
Moreover, while $\Gamma_T/U$ varies only little, $\Gamma_S/U$ 
decreases from $0.1$ at $z=250$~pm to $0.007$ at $z=630$~pm. This is a direct 
consequence of the dehybridization upon lifting the molecule. 
We note that the $\Gamma_S/U$ found here are in agreement with values independently calculated with {\it ab initio} methods \cite{Greuling11,greuling13}. In addition, in the weak-coupling regime (the inelastic spin-flip) third-order perturbation  theory \cite{appelbaum66,*appelbaum67,ternes15} yields similar estimates of the overall $\Gamma/U$ (dominated by $\Gamma_S/U$), as  Fig.~\ref{fig5_fit-ratio} shows. 

In summary, we have shown that lifting a spin-1/2 molecule from the metal surface allowed us to tune the system continuously from the strongly coupled Kondo regime to weakly coupled spin-flip regime.  
The detailed analysis of experimental data revealed inherent difficulties to reliably obtain $T_\mathrm{K}$ when thermal and Kondo energy scales are similar. However, we show that this problem can be tackled if additional data measured in magnetic fields, when the Zeeman energy is larger than the thermal energy, is used,
even when the splitting of the zero bias anomaly is not present yet. Furthermore, our scheme allows us to separate the coupling strengths to the two electron leads in asymmetrically coupled Kondo systems.

Note that Kondo temperature is the principal property of adsorbed open-shell molecules in which Kondo screening is active. Its value determines the resulting functionality of the open-shell adsorbate. For instance, if the molecular magnetic moment needs to be stabilized, the respective $T_\mathrm{K}$ must be decreased below $T_\mathrm{exp}$. On the other hand, if spin degeneracy is not wanted, $T_\mathrm{K}$ must be increased. Therefore, the unambiguous determination of $T_\mathrm{K}$ paves the way to achieving control over spin degrees of freedom at the nanoscale. 

\begin{acknowledgement}

T.~N.~and M.~\v{Z}.\ acknowledge support by the Czech Science Foundation via Project
No.~16-19640S and M.~S. via No. 17-24210Y, P.~J. was supported by Praemium Academie of the  Czech Academy of Sciences and GACR 20-13692X. We acknowledge CzechNanoLab Research Infrastructure supported by MEYS CR (LM2018110). M.~T.\ was supported by the Heisenberg Program (Grant No. TE 833/2-1)  of the Deutsche Forschungsgemeinschaft, and R.~K.\ from the PRIMUS/Sci/09 programme of the Charles University.  R.T. acknowledges support from the Young Investigator Group program (Grant No. VH-NG-514) of the Helmholtz Association. The authors thank M.~Thoss, B.~Lechtenberg, F.~B.~Anders and D.~M.~Fugger for helpful discussions. 
\end{acknowledgement}

Supporting Information: details on the sample preparation, theoretical model, the fitting procedures and analysis of additional features of the experimental $dI/dV$. SI also includes additional Refs.~\cite{Schiller_2000,ZitkoPruschke_2009,Weichselbaum07,Campo_2005,Zitko_2009,Wolf69,Brandbyge_2005,Ho_2016,Brandbyge_2006,Rakhmilevitch2014}


\providecommand{\latin}[1]{#1}
\makeatletter
\providecommand{\doi}
{\begingroup\let\do\@makeother\dospecials
	\catcode`\{=1 \catcode`\}=2 \doi@aux}
\providecommand{\doi@aux}[1]{\endgroup\texttt{#1}}
\makeatother
\providecommand*\mcitethebibliography{\thebibliography}
\csname @ifundefined\endcsname{endmcitethebibliography}
{\let\endmcitethebibliography\endthebibliography}{}

\includepdf[pages=-]{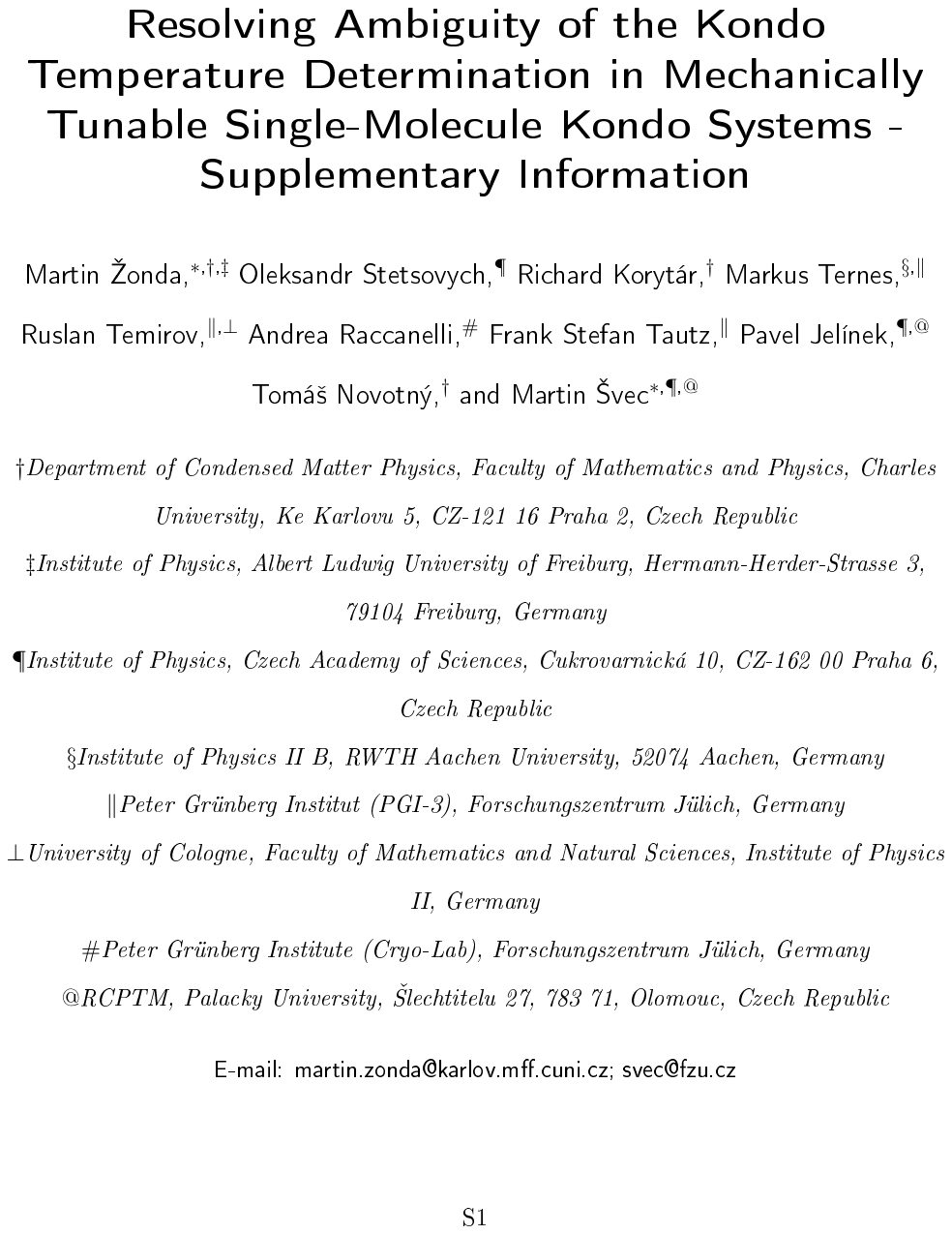}

\end{document}